\documentclass[prl,twocolumn,amsmath,amssymb]{revtex4-2}

\usepackage{graphicx}
\usepackage{dcolumn}
\usepackage{bm}
\usepackage{mathrsfs}
\usepackage{subfigure}
\usepackage{natbib}
\usepackage{amsmath}
\usepackage{amssymb}
\usepackage{Jonasmacros}
\usepackage{mathptmx}
\usepackage{txfonts}
\usepackage{pifont}

\usepackage[colorlinks,citecolor=blue,linkcolor=Fuchsia]{hyperref}
\usepackage[usenames,dvipsnames,svgnames]{xcolor}

\begin{document}

\title{Charge Redistribution and Spin Polarization Driven by Correlation Induced Electron Exchange in Chiral Molecules}

\author{J. Fransson}
\email{Jonas.Fransson@physics.uu.se}
\affiliation{Department of Physics and Astronomy, Box 516, 75121, Uppsala University, Uppsala, Sweden}




\begin{abstract}
Chiral induced spin selectivity is a phenomenon that has been attributed to chirality, spin-orbit interactions, and non-equilibrium conditions, while the role of electron exchange and correlations have been investigated only marginally until very recently. However, as recent experiments show that chiral molecules acquire a finite spin-polarization merely by being in contact with a metallic surface, these results suggest that electron correlations play a more crucial role for the emergence of the phenomenon than previously thought. Here, it is demonstrated that molecular vibrations give rise to molecular charge redistribution and accompanied spin-polarization when coupling a chiral molecule to a non-magnetic metal. It is, moreover, shown that enantiomer separation, due to spin-polarization intimately related to the chirality, can be understood in terms of the proposed model.
\end{abstract}
\maketitle

Since its discovery, chiral induced spin selectivity\cite{Science.283.814,Science.331.894} has been considered to emerge from the the combination of structural chirality, spin-orbit interactions, and strongly non-equilibrium conditions. While chirality is a prerequisite, spin-orbit interactions is suggested to be one of the cornerstones in any theoretically comprehensible description \cite{JChemPhys.131.014707,EPL.99.17006,JPCM.26.015008,PhysRevB.88.165409,JChemPhys.142.194308,PhysRevE.98.052221,PhysRevB.99.024418,NJP.20.043055,JPhysChemC.123.17043,PhysRevB.85.081404(R),PhysRevLett.108.218102,PNAS.111.11658,JPhysChemC.117.13730,PhysRevB.93.075407,PhysRevB.93.155436,ChemPhys.477.61,NanoLett.19.5253,JPhysChemLett.9.5453,JPhysChemLett.9.5753,JChemTheoryComput.16.2914,JPhysChemLett.10.7126,CommunPhys.3.178,NewJPhys.22.113023,PhysRevB.102.035431,PhysRevB.102.214303,PhysRevB.102.235416}.
Non-equilibrium conditions, arising from the probing techniques used in the measurements, for instance, light exposure \cite{Science.283.814,Science.331.894,PNAS.110.14872,NanoLett.14.6042,NatComms.7.10744,AdvMat.30.1707390,JPhysChemLett.9.2025}, local probing techniques \cite{NanoLett.11.4652,JPhysChemLett.11.1550,AdvMater.28.1957,ACSNano.14.16624}, transport \cite{NatComms.4.2256,JPhysChemLett.10.1139,JPhysChemLett.11.1550} and different types of Hall measurements \cite{NatComms.7.10744,NatComms.8.14567,AdvMat.30.1707390}, however, have typically not been regarded as part of the phenomenology. Particularly, in many theoretical considerations, non-equilibrium conditions have not been accounted for. Instead, the focus has lied on the transmission properties of chiral molecules embedded in a given environment \cite{JChemPhys.131.014707,EPL.99.17006,JPCM.26.015008,PhysRevB.88.165409,JChemPhys.142.194308,PhysRevE.98.052221,PhysRevB.99.024418,NJP.20.043055,JPhysChemC.123.17043,PhysRevB.85.081404(R),PhysRevLett.108.218102,JPhysChemC.117.13730,PhysRevB.93.075407,PhysRevB.93.155436,ChemPhys.477.61,JPhysChemLett.9.5453,JPhysChemLett.9.5753,JChemTheoryComput.16.2914,CommunPhys.3.178,NewJPhys.22.113023}. While the transmission pertains to the linear response regime, it is typically the result of a single particle description which, therefore, is not capable of resolving the chemistry or physics the molecule is subject to under non-equilibrium conditions.

In chemistry, it is well known that addition or subtraction of one or several electrons can completely change the properties of the molecule. For instance, charge, as well as, spin polarization resulting from changing the number of electrons on the molecule may vary its intrinsic properties to a degree which can only be addressed in terms of sophisticated theoretical methods.
Questions related to such structural changes were recently addressed,\cite{JPhysChemLett.10.7126,NanoLett.20.7077} stressing the vital role of electronic Coulomb interactions in this context. It was shown that Coulomb interactions generate the exchange necessary for producing measurable effects regarding, for example, chiral induced spin selectivity and enantiomer separation. Other attempts along these line, however, through electron-vibration \cite{PhysRevB.102.035431,PhysRevB.102.235416}, and polarons \cite{PhysRevB.102.214303}, have shown on the importance of expanding the theoretical concepts to more elaborate models.

In this context, it is also natural to question whether the chiral molecules maintain their intrinsically spin-degenerate properties when attached to metals. Indeed, recent experiments suggest that a strong spin-polarization can be associated with the interface between chiral molecules and metallic surfaces \cite{NatComms.8.14567,Molecules.25.6036,PNAS.114.2474,NanoLett.19.5167}. In Refs. \citenum{NatComms.8.14567,Molecules.25.6036}, chiral molecules were used to control the magnetism in a thin Co layer, resolved through the anomalous Hall effect, while enantiomer separation was concluded viable on non-magnetic metals \cite{PNAS.114.2474}, whereas Yu-Shiba-Rusinov states \cite{ProgTheorPhys.40.435,ActaPhysSin.21.75,JETPLett.2.85} were observed in the vicinity of chiral molecules on the surface of superconducting NbSe$_2$.
\cite{NanoLett.19.5167} Since the observation of Yu-Shiba-Rusinov states is strongly related with the presence of localized magnetic moments, these results vividly suggest the emergence of finite spin moments when interfacing chiral molecules with metals.
Related to these observations are also the results showing strongly enantiomer dependent binding energies on ferromagnetic metals \cite{Science.360.1331,ApplPhysLett.115.133701,JChemPhysB.123.9443,AdvMater.31.1904206}. Enantiomer separation was addressed theoretically in Ref. \citenum{NanoLett.20.7077} for molecules in contact with ferromagnetic metal, based on a description in which the electronic exchange plays a crucial role in the magnetic response. On the other hand, since these studies were made solely for molecules in an environment where the ferromagnet generates the symmetry breaking to which the electronic properties of the molecule respond, the question whether chiral molecules themselves may generate a finite spin-polarization when in contact with a metal remains open.

\begin{figure}[t]
\begin{center}
\includegraphics[width=\columnwidth]{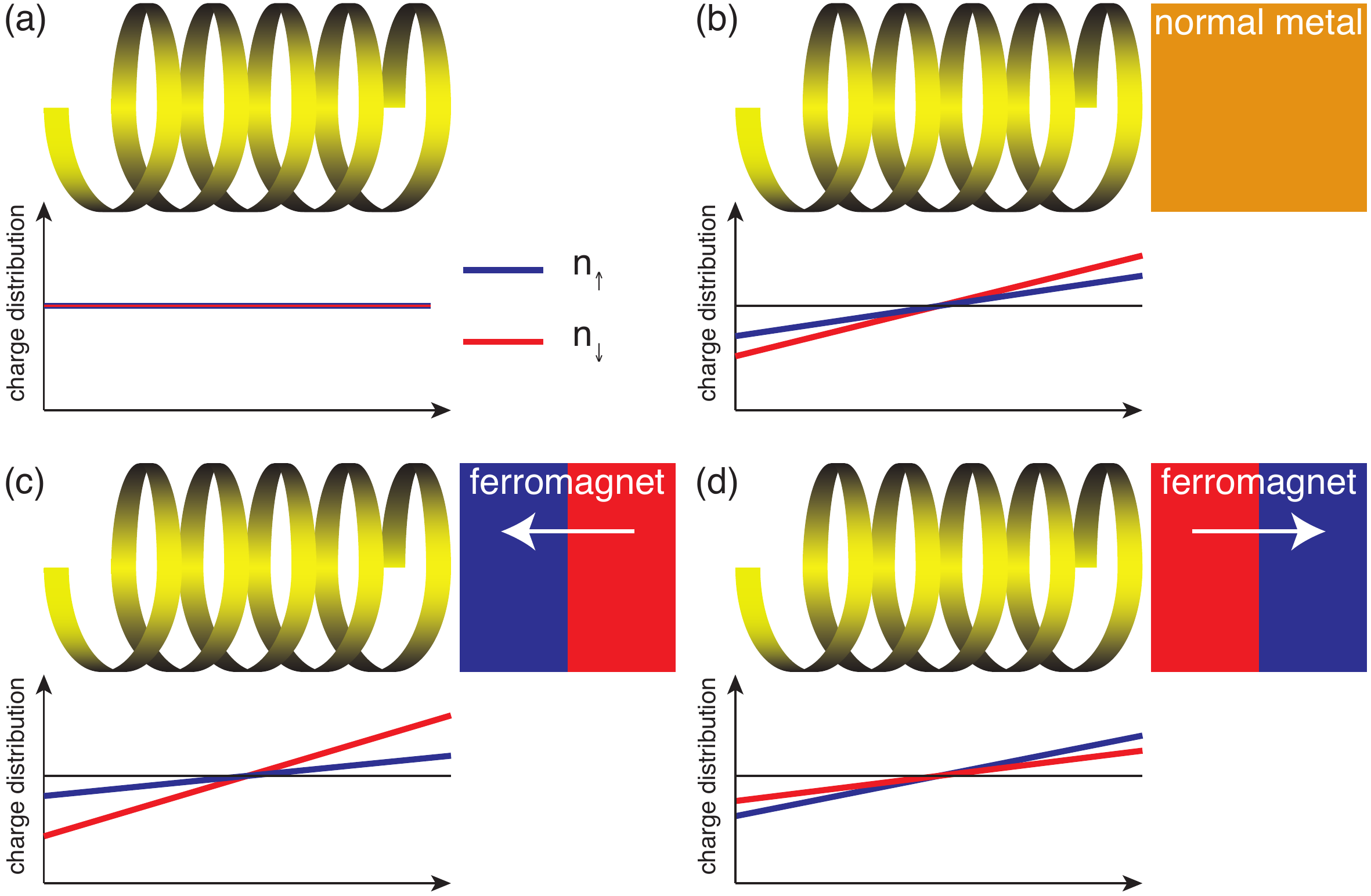}
\end{center}
\caption{Chiral molecule in (a) vacuum, (b) in contact with non-magnetic metal, (c, d) in contact with ferromagnetic metals with opposite orientations. The diagrams illustrate the spin resolved molecular charge distributions, $n_\up$ (blue) and $n_\down$ (red). The black lines indicate the charge distribution in vacuum.
}
\label{Fig1}
\end{figure}

A summary of the collected research, thus far, concerning the magnetic properties of chiral molecules is illustrated in Figure \ref{Fig1}. In vacuum, the spin-degenerate charge is uniformly distributed in the molecule (Figure \ref{Fig1}(a)). Upon coupling the molecule to a non-magnetic metal, the charge is strongly redistributed resulting in a non-vanishing charge polarization (Figure \ref{Fig1}(b)). By the chirality, the charge polarization is accompanied by a spin-polarization (Figure \ref{Fig1}(b)). The intrinsic preference of the spin-polarization can be amplified or reduced with a ferromagnet (Figure \ref{Fig1}(c, d)).

In this Letter, it is demonstrated that electron correlations originating in molecular vibrations is of crucial importance for the emergence of a finite spin-polarization in chiral molecules coupled to a metal. While the molecular structure is non-spin-polarized in vacuum, vibrationally assisted charge redistribution created in molecules attached to a metal, generates a finite spin-polarization due to chiral induced charge-spin separation. The phenomenology is unequivocally shown to be associated with molecular vibrations combined with a strongly asymmetric charge polarization. Implementation of these results in ferromagnetic environment corroborate, moreover, the importance of electron correlations as a source of an exchange splitting between the spin channels.

The discussion presented here is based on simulations of idealized chiral models of realistic, e.g., $\alpha$-helix oligopeptides and polyalanines. Since the focus lies on the cooperation between chirality, spin-orbit interactions, and electron correlations, the mapping onto specific molecular compounds is less important as the details strongly vary between different structures. The model was proposed in Ref. \citenum{PhysRevB.102.235416}. Using experimentally viable spin-orbit interaction parameters, it was shown under \emph{non-equilibrium} conditions that, the exchange splitting between the spin channels that was introduced by vibrations, supports a chiral induced spin selectivity of tens of percents. The exchange splitting is, hence, a source for a substantial non-equivalence between the spin channels, a non-equivalence which is maintained under reversal of the magnetic environment.

A fundamental difference from essentially all previous theoretical studies is that, here the molecule is attached to a single metal and no external forces are applied. Hence, the molecule establishes a (quasi-) equilibrium state with the metal, in which no charge currents flow, which is the state considered here. While rapid transient evolution is fundamentally interesting in this context, it is beyond the scope of the present discussion.
In this sense, the theory presented here is more directly appropriate for comprehending the results in, e.g., Refs. \citenum{NatComms.8.14567,Molecules.25.6036}, while the connection is more losly qualitative with the experiments reported in, e.g, Refs. \citenum{Science.360.1331,ApplPhysLett.115.133701,JChemPhysB.123.9443,AdvMater.31.1904206}.

The simulations are performed on a model of a chiral structure which constitutes a set of $\mathbb{M}=M\times N$ ionic coordinates $\bfr_m=(a\cos\varphi_m,a\sin\varphi_m,c_m)$, $\varphi_m=(m-1)2\pi/(\mathbb{M}-1)$, and $c_m=(m-1)c/(\mathbb{M}-1)$, where $a$ and $c$ define the radius and length, respectively, of the helical structure of $M$ laps with $N$ ions per lap. A Hamiltonian model can be written as
\begin{align}
\Hamil_\text{mol}=&
	\sum_{m=1}^\mathbb{M}
	\biggl(
		\dote{m}\psi^\dagger_m\psi_m
			+
		\sum_\nu
			\omega_{m\nu}b^\dagger_{m\nu}b_{m\nu}
	\biggr)
\nonumber\\&
	-
	\sum_{m=1}^{\mathbb{M}-1}
	\Bigl(
		\psi^\dagger_m\psi_{m+1}
		+
		H.c.)
	\Bigr)
	\biggl(
		t_0
		+
		\sum_\nu
			t_{m\nu}\Bigl(b_{m\nu}+b^\dagger_{m\nu}\Bigr)
	\biggr)
\nonumber\\&
	+
	\sum_{m=1}^{\mathbb{M}-2}
	\Bigl(
		i\psi^\dagger_m\bfv_m^{(+)}\cdot\bfsigma\psi_{m+2}
		+
		H.c.)
	\Bigr)
	\biggl(
		\lambda_0
		+
		\sum_\nu
			\lambda_{m\nu}\Bigl(b_{m\nu}+b^\dagger_{m\nu}\Bigr)
	\biggr)
	.
\end{align}
Here, the molecule is described by a set of single-electron energy levels $\{\dote{m}\}$, where $\dote{m}$ denotes the energy level at the position $\bfr_m$, associated with the electron creation and annihilation spinors $\psi_m^\dagger$ and $\psi_m$, respectively. Nearest-neighboring sites interact, second line in $\Hamil_\text{mol}$, via direct hopping, rate $t_0$, and electron-phonon assisted hopping, rate $t_{m\nu}$. Similarly, the spin-orbit coupling is picked up between next-nearest neighbor sites, last line in $\Hamil_\text{mol}$, through processes of the type $i\psi_m^\dagger\bfv_m^{(s)}\cdot\bfsigma\psi_{m+2s}$, $s=\pm1$, where $\lambda_0$ and $\lambda_{m\nu}$ denote the direct and electron-phonon assisted spin-orbit interaction parameters, respectively, and where $\bfsigma$ denotes the vector of Pauli matrices. The vector $\bfv_m^{(s)}=\hat\bfd_{m+s}\times\hat\bfd_{m+2s}$ defines the chirality of the helical molecule in terms of the unit vectors $\hat{\bfd}_{m+s}=(\bfr_m-\bfr_{m+s})/|\bfr_m-\bfr_{m+s}|$, and different enantiomers are, here, represented by different signs ($\pm)$ of the chirality. The electrons are at each site coupled to the vibrational modes $\omega_{m\nu}$, which are represented by the phonon operators $b_{m\nu}$ and $b^\dagger_{m\nu}$, through the rates $t_{m\nu}$ and $\lambda_{m\nu}$. For simplicity and without loss of generality, each site is modeled to carry a single vibrational mode that couples to the electronic structure on-site only. By omitting inter-site couplings, it is justified to assume that the ions vibrate with the same energy $\omega_0$ and that the on-site electron-phonon coupling parameters $t_{m\nu}=t_1$ and $\lambda_{m\nu}=\lambda_1$, for all $m$ and $\nu$.
While the spin-independent coupling $t_1$ is not strictly necessary to obtain results that are qualitatively similar to the one presented in the following discussion, it has been included since it is likely to be larger than the spin-dependent coupling $\lambda_1$. Any Coulomb repulsion has been excluded since the vibrationally induced exchange is expected to be dominating at room temperature, which is of main interest here.

The presence and properties of the metallic surface is captured by the parameter $\bfGamma=\Gamma_0(\sigma^0+p\sigma^z)/2$, which represents the coupling between the ionic site $m=1$ and the itinerant electrons in the surface. Here, $\Gamma_0=2\pi\sum_{\bfk\sigma}|v_{\bfk\sigma}|^2\rho_\sigma(\dote{\bfk})$ accounts for the spin-dependent hybridization rate $v_{\bfk\sigma}$ and spin-density of electron states $\rho_\sigma(\dote{\bfk})$ in the metal, whereas $|p|\leq1$ denotes the effective spin-polarization of the coupling.

The properties of the electronic structure are related to the single electron Green function $\bfG_{mn}(z)=\av{\inner{\psi_m}{\psi^\dagger_n}}(z)$, letting $\bfG_m\equiv\bfG_{mm}$, through, e.g., the density of electron states $\rho_m(\omega)=i{\rm sp}[\bfG^>_m(\omega)-\bfG^<_m(\omega)]/2\pi$ and spin resolved charges $\av{n_{m\sigma}}=(-i){\rm sp}(\sigma^0+\sigma^z_{\sigma\sigma}\sigma^z)\int\bfG^<_m(\omega)d\omega/4\pi$, where $\bfG^{<(>)}_m$ is proportional to the density of occupied (unoccupied) electron states. Here, ${\rm sp}$ denotes the trace over spin $1/2$ space.

The equation of motion for the Green function $\bfG_{mn}=\bfG_{mn}(z)$ can be written on the form
\begin{align}
\Bigl(
	z
	-&
	E_m
	\Bigr)
	\bfG_{mn}
	-
	\sum_{s=\pm1}
		\Biggl\{
			-t_0\bfG_{m+sn}
			+i\lambda_0\bfv_m^{(s)}\cdot\bfsigma\bfG_{m+2sn}
\nonumber\\&
			+
			\sum_{s'=\pm1}
				\bfSigma_m
				\biggl(
					t_1^2\bfG_{m+s+s'n}
					-\lambda_1^2\bfv_m^{(s)}\cdot\bfsigma\bfv_{m+2s}^{(s')}\cdot\bfsigma\bfG_{m+2(s+s')n}
\nonumber\\&
					-it_1\lambda_1\bfsigma\cdot
					\Bigl(
						\bfv_m^{(s)}\bfG_{m+2s+s'n}
						+
						\bfv_{m+s}^{(s')}\bfG_{m+s+2s'n}
					\Bigr)
				\biggr)
		\Biggr\}
	=
	\delta_{mn}
	.
\end{align}
Here, $E_1=\dote{1}-i\bfGamma/2$, which includes the level broadening due to the coupling $\bfGamma$ of this site to the metal, wheareas $E_m=\dote{m}$, $2\leq m\leq\mathbb{M}$, and $\bfG_{mn}=0$ for $m,n\notin\{1,2,\ldots,\mathbb{M}\}$. The self-energy $\bfSigma_m=\bfSigma_m(z)$ represents electron-phonon interaction loop described by
\begin{align}
\bfSigma_m(z)=&
	-\frac{1}{\beta}
	\sum_\mu
		\bfG_m(z_\mu)D_m(z-z_\mu)
	,
\end{align}
where $z_\mu=i(2n+1)\pi/\beta$ is the Fermionic Matsubara frequency, whereas $\beta=1/k_BT$ is the inverse temperature $T$ in terms of the Boltzmann constant $k_B$. In terms of the simplest non-trivial electron-phonon interactions, the self-energy is given by
\begin{align}
\bfSigma_m(z)=&
			\frac{n_B(\omega_0)+1-f(\dote{m})}{z-\omega_0-\dote{m}}
			+
			\frac{n_B(\omega_0)+f(\dote{m})}{z+\omega_0-\dote{m}}
	,
\label{eq-Sigma}
\end{align}
where $n_B(\omega)$ and $f(\omega)$ are the Bose-Einstein and Fermi-Dirac distribution functions, respectively. Insofar the exchange splitting generated by the molecular vibrations can be regarded as an intrinsic property of the structure, the employed approach is justified since it captures the main effect of the electron-phonon coupling. Hence, despite the charge redistribution may modify the exchange splitting, the gross effect of the electron-phonon interactions is captured by using the approximation of instantaneous thermalization of the vibrations.

\begin{figure}[t]
\begin{center}
\includegraphics[width=\columnwidth]{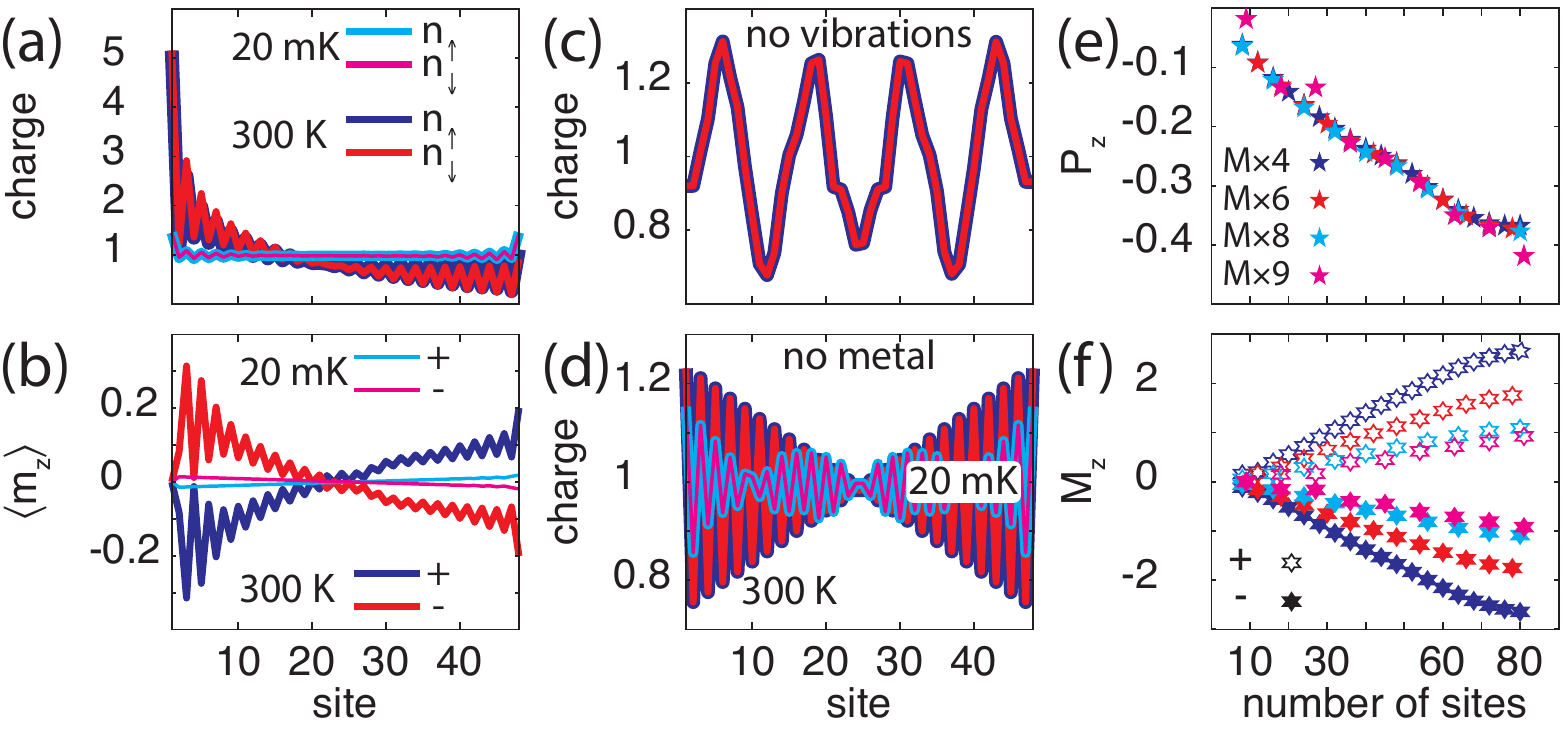}
\end{center}
\caption{Chiral molecule ($8\times6$) in contact with a metallic surface.
(a, b) Spin-resolved charge distribution and corresponding spin-polarization ($\av{m_z}=(n_\up-n_\down)/2$) per site in the chiral molecule at the temperatures $T=20$ mK and 300 K. In (a), the blue (cyan) and red (magenta) lines correspond to $n_\up$ and $n_\down$, respectively, at 300 K (20 mK). In (b), the blue (cyan) and red (magenta) lines represent positive ($+$) and negative ($-$) helicity, respectively, at 300 K (20 mK).
(c) Spin-resolved charge distribution per site in the static chiral molecule at 20 mK (cyan, magenta) and 300 K (blue, red).
(d) Spin-resolved charge distribution per site in the isolated vibrating chiral molecule at 20 mK (cyan, magenta) and 300 K (blue, red).
(e, f) Charge and spin polarizations $P_z$ and $M_z$, respectively, as function of the number of sites, at 300 K. (f) Positive (negative) helicity is denoted by open (filled) hexagons.
Here, parameters used in the simulations are, in units of $t_0=40$ meV: $\dote{0}-E_F=-2$, $\Gamma_0=1/10$, $\omega_0=1/100$, $\lambda_0=1/40$, $t_1=1/10$, and $\lambda_1=1/400$, where $E_F$ is the Fermi energy of the metal.
An intrinsic broadening $1/\tau_\text{ph}=t/4$ was used in the vibration self-energy in order to smoothen the electronic densities.
In panel (d), the long wave length substructure is attributed to the numerical sensitivity to the integration mesh at low temperature.
}
\label{Fig2}
\end{figure}

The plots in Figure \ref{Fig2}(a) show the charge distribution for the vibrating molecule mounted on the metallic surface. The charge distribution is at 300 K (i) strongly redistributed, with depleted charge in the interior of the molecule accumulating near the metal, and (ii) accompanied by a non-vanishing spin-polarization, see Figure \ref{Fig2}(a, b). At low temperatures (20 mK), the charge is strongly confined to its bare electronic structure, since the vibrational excitations are thermally suppressed, which leads to a substantially weakened charge redistribution. Nevertheless, the small amount of charge reorganization that does occur is also accompanied by a non-vanishing spin-polarization, albeit much weaker than at elevated temperatures. The orientation of the emerging spin-polarization depends on the chirality of the molecules, Figure \ref{Fig2}(b), which is expected since only the chirality change upon shifting helicity from positive to negative. It should be noted that the absence of spin-polarization of the site adjacent the surface is an effect of the fixed boundary conditions.
As reference these results are compared with the result of the static molecule, Figure \ref{Fig2}(c), and the vibrating molecule in vacuum, Figure \ref{Fig2}(d). In both configurations, the charge distribution is weakly non-uniform, and symmetric around the center of the molecule along its length direction, with vanishing spin-polarization, both at low and high temperatures.
The charge variations in Figure \ref{Fig2}(c) are due to the strongly delocalized nature of the electrons, such that the site index is not a good quantum number. Hence, the charge may accumulate or deplete non-uniformly throughout the structure while the total charge is conserved.

The charge polarization, $P_z=2\sum_m(c_m-\av{z})\av{n_m}/\mathbb{M}L$, $\av{z}=\sum_mc_m/\mathbb{M}$, $\av{n_m}=\sum_\sigma\av{n_{m\sigma}}$, and $L=c\mathbb{M}$, can be used for a normalized collective measure of the charge redistribution, such that $|P_z|\leq1$. In this context, then, a negative (positive) charge polarization should be understood as a charge accumulation (depletion) in the end of the molecule adjacent to the metal, and charge depletion (accumulation) in the free end of the molecule, (Figure \ref{Fig2}(a, c, d)). 
In analogy to the charge polarization, a measure of the mean spin-polarization is given by $M_z=2\sum_m(c_m-\av{z})\av{n_{m\up}-n_{m\down}}/L$. A negative (positive) value of this measure should, accordingly, be interpreted as an overweight of spin $\up$ ($\down$) near the metal and/or an overweight of the opposite spin on the free end, see Figure \ref{Fig2}(b). %
Here, $P_z$ and $M_z$ associated with vibrating molecules at 300 K are plotted in Figure \ref{Fig2}(e, f), as function of the molecule length. The plots demonstrate the growth of both charge and mean spin-polarization with molecular length. The charge polarization appears, in addition, to be a universal feature, as the values of the different types of molecules fall on essentially the same line. The deviations ($M\times9$) are an effect of the finite size which tend to vanish with increasing length. Absence of either vibrations or coupling to external environment leads to vanishing charge and mean spin-polarizations, (Figure \ref{Fig2}(c, d)).
The results summarized in Figure \ref{Fig2} confirm that charge polarization, and accompanied spin-polarization, in the composite system originates in electron correlations.
 
\begin{figure}[t]
\begin{center}
\includegraphics[width=\columnwidth]{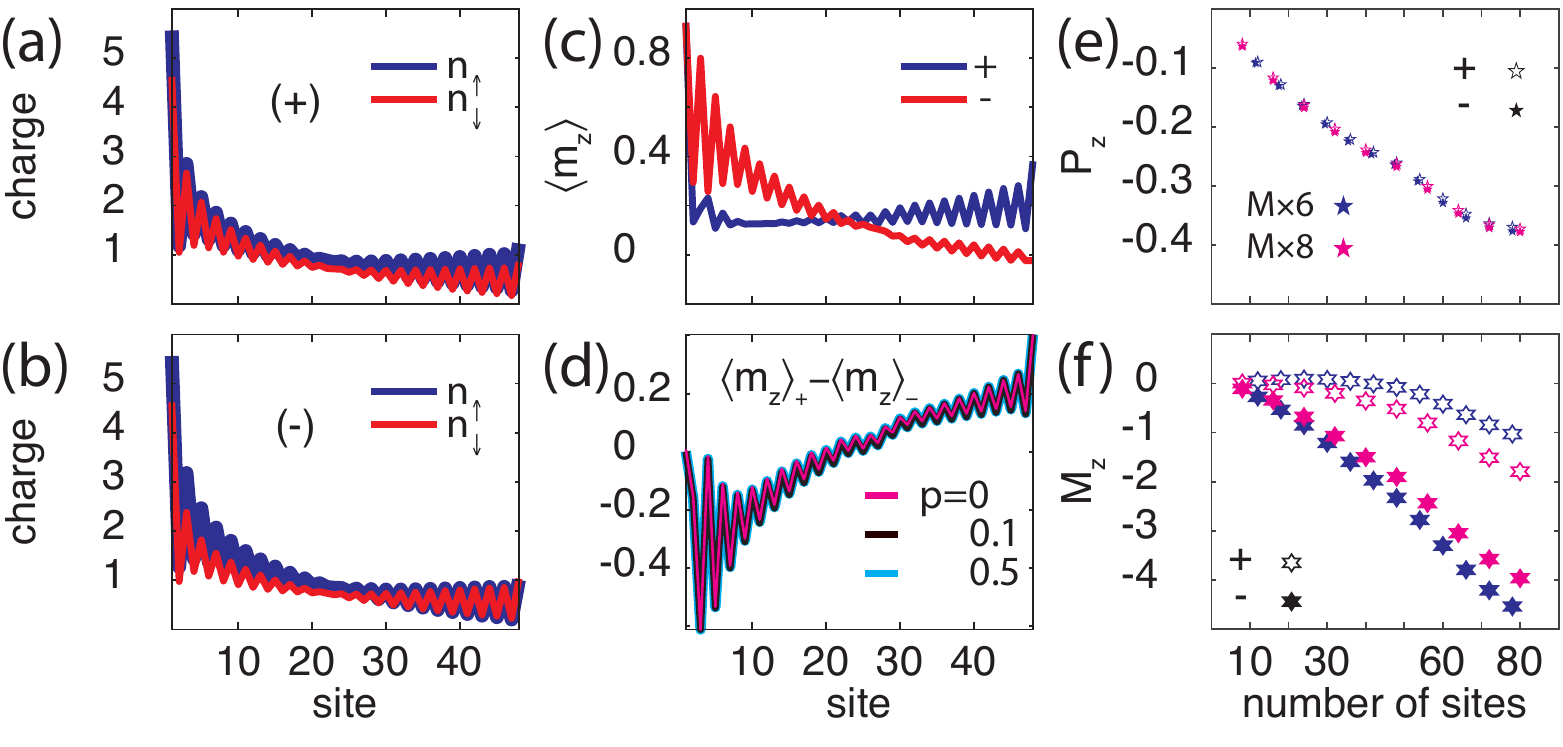}
\end{center}
\caption{Chiral molecule ($8\times6$) in contact with a ferromagnetic surface.
(a, b) Spin-resolved charge distribution for a molecule with (a) positive and (b) negative helicity.
(c) Corresponding spin-polarization per site in the chiral molecule, for (blue) positive ($+$) and (red) negative ($-$) helicity.
(d) Difference $\av{m_z}_+-\av{m_z}_-$ between the spin-polarizations for positive and negative helicity.
(e, f) Charge and spin polarizations $P_z$ and $M_z$, respectively, as function of the number of sites, for positive (open symbols) and negative (filled symbols) helicity.
Here, $p=0.1$ and $T=300$ K, while other parameters are as in Figure \ref{Fig2}.}
\label{Fig3}
\end{figure}

The above results demonstrate that a finite spin-polarization emerges in the chiral molecule when interfaced with a metal. Since the sign of the spin-polarization depends on the chirality, with an overweight of spin $\down$ ($\up$) near the interface for positive (negative) helicity, this spin-polarization is expected to be diminished (enhanced) by a positive spin-polarization ($p>0$) in the coupling parameter $\bfGamma$. This expectation is corroborated in the spin-resolved charge distributions shown in Figure \ref{Fig3}, for (a) positive and (b) negative helicity, and (c) corresponding spin-polarizations, in the set-up with $p=0.1$. Here, the spin-polarization of the molecule with positive helicity has a tendency to counteract the external spin-polarization, in a fashion which is not unlike a diamagnetic property. Opposite (negative) helicity tends, on the other hand, to magnetically act cooperatively with the external spin-polarization.
The loss of mirror symmetry between the plots in Figure \ref{Fig3} (a) and (b) is an expected outcome of the vibrationally generated exchange.
However, despite the molecular spin-polarization tends to be strongly modified by the external conditions, the intrinsic properties of the chiral molecules remain unchanged when comparing positive and negative helicity. This is shown by the difference $\av{m_z}_+-\av{m_z}_-$ (helicity $\pm$) in Figure \ref{Fig3}(d). The difference in the induced spin-polarizations is the same for any $p$, here shown for $p=0$, $0.1$, and $0.5$, something which indicates an intrinsic anisotropy.
Since a corresponding universal difference between the spin-polarizations does not exist in absence of the
vibrationally generated correlations (not shown),
this result shows a stability of the correlation induced spin-polarization.
Moreover, while an external spin-polarization $p$ tends to result in a slightly different charge polarization $P_z$ for different helicity (Figure \ref{Fig3}(e)), although minute, the mean spin-polarization $M_z$ is strongly affected (Figure \ref{Fig3}(f)), in accordance with the expected characteristics.
The length dependence of the mean spin-polarization $M_z$ is in good agreement with the results reported in, e.g., Refs. \citenum{Science.331.894,Science.360.1331}, especially for longer chains whereas the agreement is not as well for short.

\begin{figure}[t]
\begin{center}
\includegraphics[width=\columnwidth]{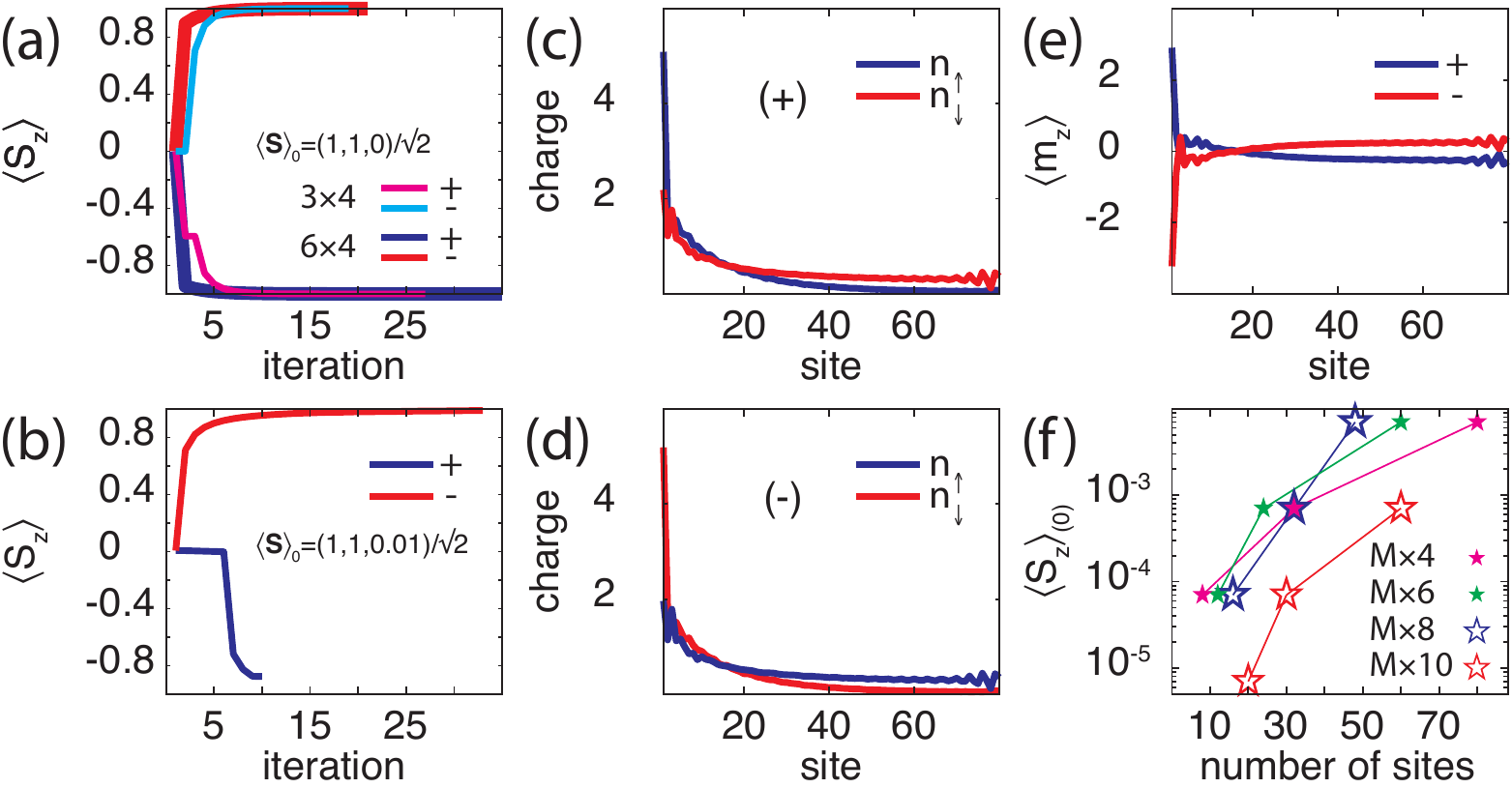}
\end{center}
\caption{
(a, b) Evolution of $\av{S_z}$ of the local spin moment $\bfS$ under the self-consistent simulations for (a) $3\times4$ and $6\times4$ molecules, and (b) $20\times4$ molecule.
(c, d) Spin-resolved molecular charge distributions in the self-consistent limit for (c) positive ($+$), and (d) negative ($-$) helicity.
(e) Spin-polarizations $\av{m_z}_\pm$ corresponding to the charge distributions in panels (c, d).
Maximum value of $\delta=\av{S_z}_0$ in the initial spin $\av{\bfS}_0=(1,1,\pm\delta)/\sqrt{2}$ that can be switched for different molecules characterized by $M\times N$.
Here, $\dote{0}-E_F=-1/2$, in panels (c,...,f) $t_1=1/20$, $\lambda_0=1/10$, $\lambda_1=1/20$, $v=1$ in units of $t_0$, while other parameters and parameters in panel (a) are as in Figure \ref{Fig2}.}
\label{Fig4}
\end{figure}

The magnetic anisotropy generated by the chiral molecule can be shown to influence an external spin moment $\bfS$ which is coupled to the molecule via exchange $v$, modelled as $v\psi^\dagger_1\bfsigma\cdot\bfS\psi_1$. Putting $p=0$, such that the coupling $\bfGamma=\Gamma_0\sigma^0/2$, ensures that the model describes a spin moment embedded in a non-magnetic metal, corresponding to the set-up in, e.g., Ref. \citenum{NatComms.8.14567}. Self-consistent calculations with respect to the molecular charge distribution show that an initial spin moment $\av{\bfS}_0=(\cos\varphi,\sin\varphi,0)/\sqrt{2}$, $0\leq\varphi<2\pi$, eventually reaches the final state $\av{\bfS}=\mp(0,0,1)$ for $\pm$ helicity (Figure \ref{Fig4}(a)). This result demonstrates the existence of an intrinsically sustained anisotropy which can be coupled to and influence the magnetic properties of the environment. Here, the local moment is represented by a spin $S=1$, for which the expectation value is provided by $\av{\bfS}=\sum_\alpha\bra{\alpha}\bfS\ket{\alpha}$, with respect to the spin Hamiltonian $\Hamil_\text{S}=iv\bfS\cdot{\rm sp}\int\bfsigma\bfG^<_1(\omega)d\omega/4\pi$.

By additionally exploring the intrinsic properties, adjusting the molecule parameters (see Figure caption for details), it can, furthermore, be demonstrated that the intrinsic anisotropy is sufficiently strong to switch the spin moment from an initial $\av{S_z}=\pm\delta$, $0\leq\delta\ll1$, to a final $\av{S_z}\mp1$ for $\pm$ helicity, which is shown in Figure \ref{Fig4}(b). The plots show the self-consistency evolution of the local moment $\av{\bfS}_0=(1,1,0.01)/\sqrt{2}$ under the influence of a $20\times4$ molecule for positive (blue) and negative (red) helicity. The corresponding spin-resolved charge distributions for the molecules are shown in Figure \ref{Fig4}(c,d), and spin-polarization $\av{m_z}$ in Figure \ref{Fig4}(e), demonstrating that the symmetry under the change of helicity is maintained also when the molecule is coupled to an external spin moment.

In the experimental observations of spin reversal using chiral molecules \cite{NatComms.8.14567,Molecules.25.6036}, the magnetization of a magnetized ferromagnetic layer, saturated in an out-of-plane configuration, was entirely switched. The simulations presented here, show a weaker anisotropy associated with the magnetic properties of the composite system. The plots in Figure \ref{Fig4}(f) display results of systematic simulations, all indicating an upper bound of $\delta$, in the initial spin moment $\av{\bfS}_0=(\cos\varphi,\sin\varphi,\pm\delta)/\sqrt{2}$. Despite this limitation of the presented theory, it, nevertheless, shows that molecular vibrations act in the composite system as to generate strong magnetic anisotropies on externally located spin moments.

In summary, it has been shown that molecular vibrations in composite molecule-metal configurations is a mechanism that breaks the spin symmetry of the molecule, in accordance with experimental observations. While the vibrational source of exchange is particularly effective at high temperatures, it is non-negligible also at lower. It was, moreover, shown that this mechanism provides an origin for enantiomer separation using magnetic measurements.
Temperature dependent anomalous Hall measurements may provide evidence for the vibrationally induced exchange.


The author thanks R. Naaman for constructive and encouraging discussions.
Support from Vetenskapsr\aa det, Stiftelsen Olle Engkvist Byggm\"astare is acknowledged.

\end{document}